\documentclass[twocolumn,eqsecnum,prb]{revtex4}
\usepackage{amsfonts}
\usepackage[T1]{fontenc}
\usepackage{amsmath,amsbsy,amssymb,graphicx}
\usepackage{times}

\def\beginABC{\begin{subequations}}
\def\endABC{\end{subequations}}
\let\mathbf=\boldsymbol

\begin{document}

\title{{\Large Monolayer Topological Insulators: Silicene, Germanene and
Stanene }}
\author{Motohiko Ezawa}
\affiliation{Department of Applied Physics, University of Tokyo, Hongo 7-3-1, 113-8656,
Japan }

\begin{abstract}
We report the recent progress on the theoretical aspects of monolayer
topological insulators including silicene, germanene and stanene, 
which are monolayer honeycomb structures of silicon, germanium and tin, respectively.
They show quantum spin Hall effects in nature due to the spin orbit interaction. 
The band gap can be tuned by applying perpendicular electric field, 
which induces a topological phase transition. 
We also analyze the topological properties of generic honeycomb systems together with the classification of topological insulators.
Phase diagram of topological insulators and superconductors in honeycomb systems are explicitly determined.
We also investigate topological electronics including a topological field-effect transistor, the topological Kirchhoff's law and the topological spin-valleytronics.
\end{abstract}

\maketitle


\section{Introduction}

Monolayer materials are one of the most active fields of condensed matter
physics. Graphene, monolayer honeycomb structure of carbon atoms, is the
first experimentally realized monolayer material. Its low-energy band
structure is described by the Dirac theory, which results in various novel
physical properties\cite{GrapheneRMP,DasRMP,Goerbig,Kotov,KatsText}. The
success of graphene evokes an extensive search for other monolayer
materials. In particular, monolayer topological materials are fascinating,
realizing topological insulators and topological superconductors.

A natural question is whether other monolayer honeycomb systems purely made
of one kind of atoms are possible. It is shown that monolayer honeycomb
systems made of silicon, germanium and tin are possible, which are named
silicene, germanene and stanene, respectively. Silicene is named after the
combination of silicon and suffix "ene", which means the sp$_{2}$ bonding
structure\cite{Guzman}. Germanene is also named after germanium $+$ "ene".
Stanene is named after the Latin word "stannum" for tin\cite{Stanene}.

Silicene, Germanene and stanene are expected to be topological insulators.
Topological insulator (TI) is a distinctive state of matter indexed by
topological numbers, and characterized by an insulating gap in the bulk
accompanied by topologically protected gapless edges\cite{Hasan,Qi}. Thus
the physics of these materials is located at the confluence of graphene and
topological insulators, which results in very rich physics.

Silicene and germanene are proposed by first-principles calculations\cite%
{Shiraishi,Guzman,Ciraci,Lebegue}, where their stability and the emergence
of the Dirac cone are predicted by first-principles calculations. It is
shown that the band gap of silicene is electrically controllable by applying
perpendicular electric field to silicene plane by the first-principle
calculations\cite{Falko,NL} and based on the Dirac theory\cite{EzawaNJP}. A
first-principles calculation shows that the Dirac cones are hidden in
silicene fabricated on the Ag substrate due to the strong hybridization
between silicene and Ag substrate\cite{Oshiyama}.

First suggestive observations\cite{Lalmi,Padova,Aufray} of silicene were
reported in 2010. Silicene has been grown on various substrates such as the
Ag substrate\cite{Lay,Takagi,Wu} and the ZrB$_{2}$ substrate\cite{Takamura},
the Ir substrate\cite{Ir} and the MoS$_{2}$ substrate\cite{MoS}. Germanene
is synthesized on the Au substrate\cite{GerLay} and Pt substrate\cite{GerPt}. 
There are several experiments on silicene\cite%
{LL,Fukaya,40,Takahashi,Hasegawa}. In particular, silicene was demonstrated
in 2015 to act as a field-effect transistor at room temperature\cite{Tao}.
There are reviews on experimental aspects of silicene\cite%
{KaraRev,TangRev,TakamuraRev}.

There are several proposal on realizing free-standing like silicene on
substrates by first-principles calculations. Silicene will be grown on
graphene\cite{Cai}, hexagonal boron-nitride\cite{BN}, hydrogen-processed
Si(111) surface\cite{Oshiyama}, hydrogen-processed Ge(111) surface\cite{Koko}, 
Cl-passivated Si(111) and clean CaF$_2$(111) surfaces\cite{KokoJ}, solid
argon\cite{Argon} , between bilayer graphene\cite{Amal} and intercalating
alkali metal atoms between silicene and the metal substrates\cite{Quhe}.

\section{Graphene and Silicene}

\begin{figure}[t]
\centerline{\includegraphics[width=0.4\textwidth]{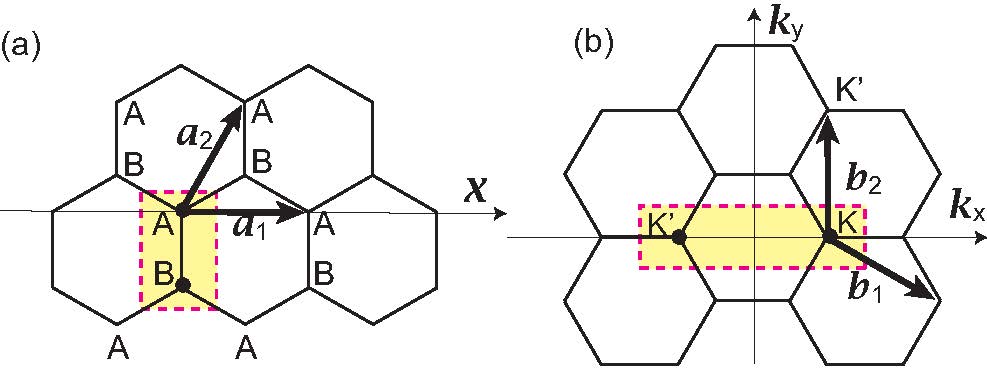}}
\caption{(a) The honeycomb structure, made of two fundamental vectors 
$\mathbf{a}_{1}$ and $\mathbf{a}_{2}$, consists of two sublattices made of $A$
and $B$ sites. A dotted rectangular represents a unit cell. (b) The
reciplocal lattice is also a honeycomb lattice. A dotted rectangular
represents a unit cell, which contains two inequivalent points $K$ and $K^{\prime }$. }
\label{FigABsite}
\end{figure}

The basic structure of graphene and silicene is a honeycomb lattice
generated by the fundamental translational vectors $\mathbf{a}_{1}$ and $\mathbf{a}_{2}$. 
It consists of two triangular sublattices made of
inequivalent lattice sites $A$ and $B$ [Fig.\ref{FigABsite}(a)]. The
reciprocal lattice is also a honeycomb lattice in the momentum space [Fig.\ref{FigABsite}(b)], which constitutes the Brillouin zone.

\subsection{Graphene}

\begin{figure}[t]
\centerline{\includegraphics[width=0.4\textwidth]{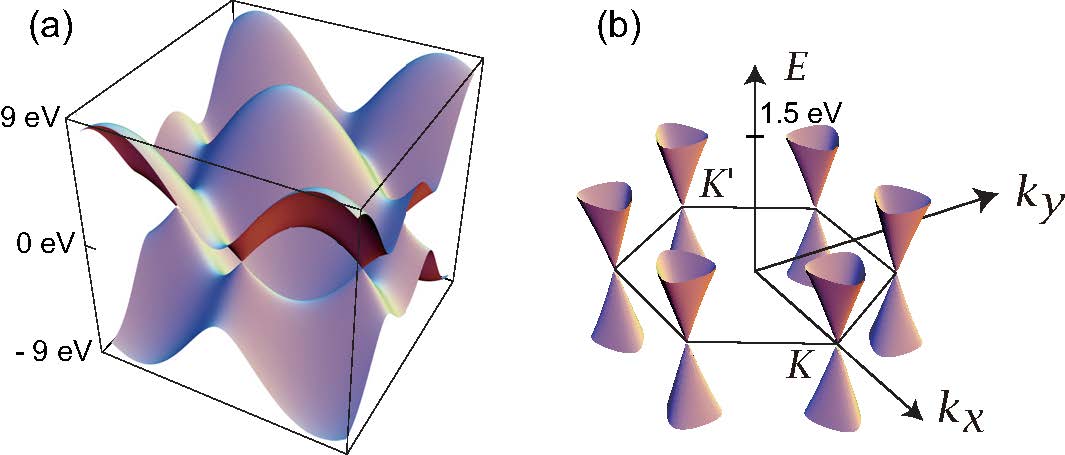}}
\caption{Band structure of graphene. (a) Six valleys are seen in this
figure. (b) The gap is closed at the $K$ and $K^{\prime }$ points, where the
band structure looks like a cone. It is called the Dirac cone because the
dispersion is linear. Among six cones only two are physically inequivalent.}
\label{FigDiracValley}
\end{figure}

Graphene is described by the simplest tight-binding model on a honeycomb
lattice [Fig.\ref{FigDiracValley}],%
\begin{equation}
\hat{H}_{0}=-t\sum_{\left\langle i,j\right\rangle s}c_{is}^{\dagger }c_{js},
\label{HamilGra}
\end{equation}%
where $c_{is}^{\dagger }$ creates an electron with spin polarization 
$s=\uparrow \downarrow $ at site $i$, $\left\langle i,j\right\rangle $ runs
over all the nearest neighbor hopping sites, and $t$ is the transfer energy.
By diagonalizing the Hamiltonian we obtain the band structure, which we
illustrate in Fig.\ref{FigDiracValley}(a). It consists of valleys or cones
near the Fermi surface. The cones touch the Fermi surface at two
inequivalent points, that is, the $K$ and $K^{\prime }$ points in the
Brillouin zone [Fig.\ref{FigDiracValley}(b)].

\begin{figure}[t]
\centerline{\includegraphics[width=0.5\textwidth]{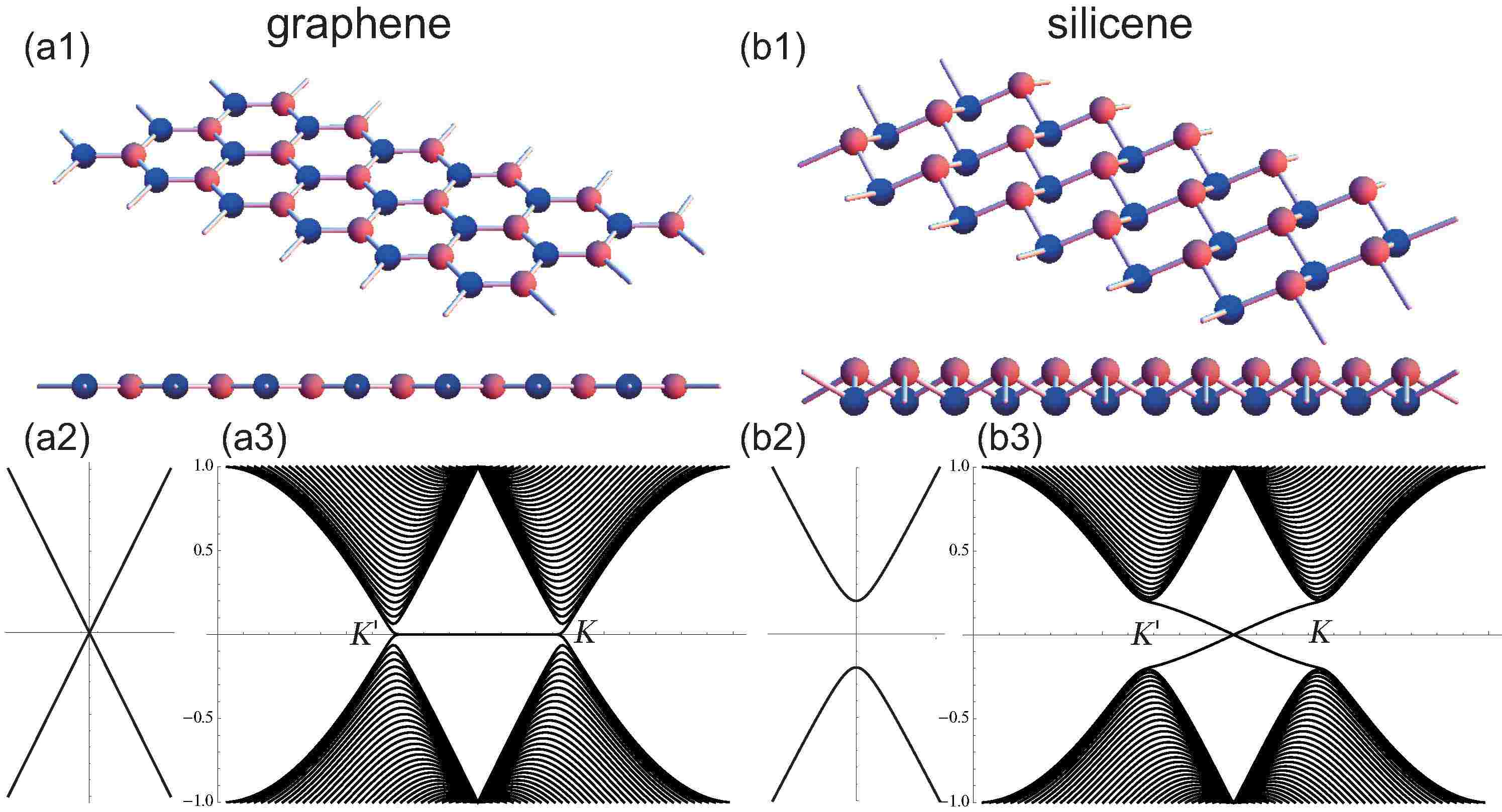}}
\caption{(a1) The lattice structure of graphene is planar, but (b1) that of
silicene is buckled. Red and blue balls represent $A$ and $B$ sites.\ (a2)\
The\ band gap of graphene is closed, where the dispersion is linear near the
Fermi energy. (b2) The band gap of silicene is open. (a3) A flat line
connecting the $K$ and $K^{\prime }$ points represents gapless flat edge
modes in a nanoribbon. It contains 4-fold degenerate edge states for
up/down-spin and left/right movers. (b3) Two lines connecting the tips of
the Dirac cones represents gapless edge modes of a nanoribbon. Each line
contains 2-fold degenerate edge states. }
\label{FigGraSili}
\end{figure}

We are interested in physics near the Fermi energy. To derive the relevant
Hamiltonian, we rewrite (\ref{HamilGra}) as%
\begin{equation}
\hat{H}_{0}=t\sum_{s}\int d^{2}k^{\prime }\left( c_{\text{A}s}^{\dagger },c_{\text{B}s}^{\dagger }\right) \left( 
\begin{array}{cc}
0 & f\left( \mathbf{k}\right) \\ 
f^{\ast }\left( \mathbf{k}\right) & 0%
\end{array}\right) \left( 
\begin{array}{c}
c_{\text{A}s} \\ 
c_{\text{B}s}%
\end{array}
\right)
\end{equation}
in the momentum space, with
\begin{equation}
f\left( \mathbf{k}\right) =e^{-iak_{y}/\sqrt{3}}+2e^{iak_{y}/2\sqrt{3}}\cos 
\frac{ak_{x}}{2}.
\end{equation}
The energy spectrum is obtained as
\begin{equation}
E\left( \mathbf{k}\right) =t\sqrt{1+4\cos \frac{ak_{x}}{2}\cos \frac{\sqrt{3}%
ak_{y}}{2}+4\cos ^{2}\frac{ak_{x}}{2}}.
\end{equation}
The gap closes at the $K_{\eta }$ point defined by
\begin{equation}
K_{\eta }=\frac{1}{a}\left( \eta \frac{4\pi }{3},0\right) \qquad \text{with}
\qquad \eta =\pm \text{.}
\end{equation}
The $K_{+}$ and $K_{-}$ points are identical to the $K$ and $K^{\prime }$
points, respectively. Because the dispersion relation is linear for $%
k_{i}\simeq 0$, they are also called the Dirac points: See Figs.\ref{FigDiracValley} and \ref{FigGraSili}(a2).

In the vicinity of the $K_{\eta }$ point, the Hamiltonian is approximated by
\begin{equation}
\hat{H}_{\eta }=\sum_{s}\int d^{2}k(c_{\text{A}s}^{\eta \dagger },c_{\text{B}%
s}^{\eta \dagger })H_{s_{z}}^{\eta }\left( 
\begin{array}{c}
c_{\text{A}s}^{\eta } \\ 
c_{\text{B}s}^{\eta }
\end{array}
\right) ,  \label{GraphHamilB}
\end{equation}
with
\begin{equation}
H_{\eta }^{s}=\hbar v_{\text{F}}\left( \eta k_{x}\tau _{x}+k_{y}\tau
_{y}\right) =\hbar v_{\text{F}}\left( 
\begin{array}{cc}
0 & \eta k_{x}-ik_{y} \\ 
\eta k_{x}+ik_{y} & 0
\end{array}
\right) ,
\end{equation}
where $\mathbf{\tau }=(\tau _{x},\tau _{y},\tau _{z})$ is the Pauli matrix
of the sublattice pseudospin for the $A$ and $B$ sites, and $v_{\text{F}}=%
\frac{\sqrt{3}}{2\hbar }at$ is the Fermi velocity with $a$\ being the
lattice constant. The dispersion relation is linear for $k_{i}\simeq 0$. We
refer to $H_{\eta }^{s}$ as the Dirac Hamiltonian at the Dirac point $K_{\eta }$. 

\subsection{Silicene and tunable band gap}

The basic nature of silicene is described also by the tight-binding model (\ref{HamilGra}). 
There are two additional features making silicene
essentially different from graphene. One is the presence of the spin-orbit
interaction, which makes silicene a topological insulator\cite{LiuPRL}. The
other is its buckled structure with a layer separation between the two
sublattices [Fig.\ref{FigGraSili}(b1)]. This freedom allows us to tune the
gap by introducing a potential difference between the two sublattices\cite%
{Falko,NL,EzawaNJP}. When we apply electric field $E_{z}$ perpendicular to
silicene, the tight-binding Hamiltonian reads
\begin{eqnarray}
\hat{H} &=&-t\sum_{\left\langle i,j\right\rangle s}c_{is}^{\dagger }c_{js}
+i\frac{\lambda _{\text{SO}}}{3\sqrt{3}}\sum_{\left\langle \!\left\langle
i,j\right\rangle \!\right\rangle s}s\nu _{ij}c_{is}^{\dagger }c_{js}  \notag
\\
&&-\ell \sum_{is}\mu _{i}E_{z}c_{is}^{\dagger }c_{is},  \label{HamilSilic}
\end{eqnarray}
where $\left\langle \!\left\langle i,j\right\rangle \!\right\rangle $ run
over all the next-nearest neighbor hopping sites. The spin index stands for $%
s=\uparrow \downarrow $ for indices and for $s=\pm $ within equations. It
describes germanene and stanene as well.

We explain each term. (i) The first term represents the usual
nearest-neighbor hopping with the transfer energy $t$. (ii) The second term
represents the effective SO coupling with $\lambda _{\text{SO}}$, where $\nu
_{ij}=+1$ if the next-nearest-neighboring hopping is anticlockwise and $\nu
_{ij}=-1$ if it is clockwise with respect to the positive $z$ axis\cite{KaneMele}. 
(iii) The third term represents the staggered sublattice
potential with $\mu _{i}=+1$ ($-1$) for the $A$ ($B$) site\cite{EzawaNJP}.
Explicit values of these parameters are summarized in the Table \ref{ParamGSGS}. 
By diagonalizing the Hamiltonian by setting $E_{z}=0$, we
obtain the band structure illustrate as in Fig.\ref{FigGraSili}(b2). The
prominent feature is that the gap is open due to the SO interaction, and
hence silicene is an insulator. A large SO interaction with $\lambda _{\text{SO}}=0.3$eV 
is materialized in functionalized stanene \cite{Stanene}, which
will be a topological insulator at room temperature.

\begin{table}[tbp]
\begin{center}
\begin{tabular}{|l|l|l|l|l|l|l|l|}
\hline
& $t$(eV) & $v$ & $a$(\r{A}$)$ & $\lambda _{\text{SO}}$ & $\lambda _{\text{R}}$ & 
$\ell $ & $\theta $ \\ \hline
Graphene & 2.8 & 9.8 & 2.46 & 10$^{-3}$ & 0 & 0 & 90 \\ \hline
Silicene & 1.6 & 5.5 & 3.86 & $3.9$ & 0.7 & 0.23 & 101.7 \\ \hline
Germanene & 1.3 & 4.6 & 4.02 & $43$ & 10.7 & 0.33 & 106.5 \\ \hline
Stanene & 1.3 & 4.9 & 4.70 & $43$ & 9.5 & 0.33 & 107.1 \\ \hline
\end{tabular}
\end{center}
\caption{The parameters charactering graphene, silicene and germanene .
Here, $v_{\text{F}}$ is in the unit of 10$^{5}$m/s, and $\protect\lambda _{\text{SO}}$ 
in the unit of meV. $\protect\lambda _{\text{R}}$ is the Rashba
SO interaction strength in the unit of meV: See (\protect\ref{Rashba}). 
$\ell $ is the buckle height, while $\protect\theta $ is the bond angle.
Taken from Ref. \protect\cite{LiuPRB}.}
\label{ParamGSGS}
\end{table}

The low-energy physic near the Fermi energy is described by the Dirac
theory, which is constructed just as in the case of graphene. We rewrite the
Hamiltonian (\ref{HamilSilic}) in the form of (\ref{GraphHamilB}). The Dirac
Hamiltonian is explicitly given by
\begin{equation}
H_{s}^{\eta }=\left( 
\begin{array}{cc}
\Delta _{s}^{\eta } & \hbar v_{\text{F}}(\eta k_{x}-ik_{y}) \\ 
\hbar v_{\text{F}}(\eta k_{x}+ik_{y}) & -\Delta _{s}^{\eta }
\end{array}
\right) ,  \label{DiracSilic}
\end{equation}
where
\begin{equation}
\Delta _{s}^{\eta }=\eta s\lambda _{\text{SO}}-\ell E_{z}\equiv -\ell
(E_{z}-\eta sE_{\text{cr}}),  \label{DiracMassEz}
\end{equation}
with
\begin{equation}
E_{\text{cr}}\equiv \lambda _{\text{SO}}/\ell .
\end{equation}
Note that $\Delta _{s}^{\eta }$ acts as the Dirac mass. The energy spectrum
reads
\begin{equation}
E\left( \mathbf{k}\right) =\pm \sqrt{\left( \hbar v_{\text{F}}k\right)
^{2}+\left( \Delta _{s}^{\eta }\right) ^{2}}.
\end{equation}
The gap is given by $2|\Delta _{s}^{\eta }|=2\ell |E_{z}-\eta sE_{\text{cr}%
}| $.

\begin{figure}[t]
\centerline{\includegraphics[width=0.23\textwidth]{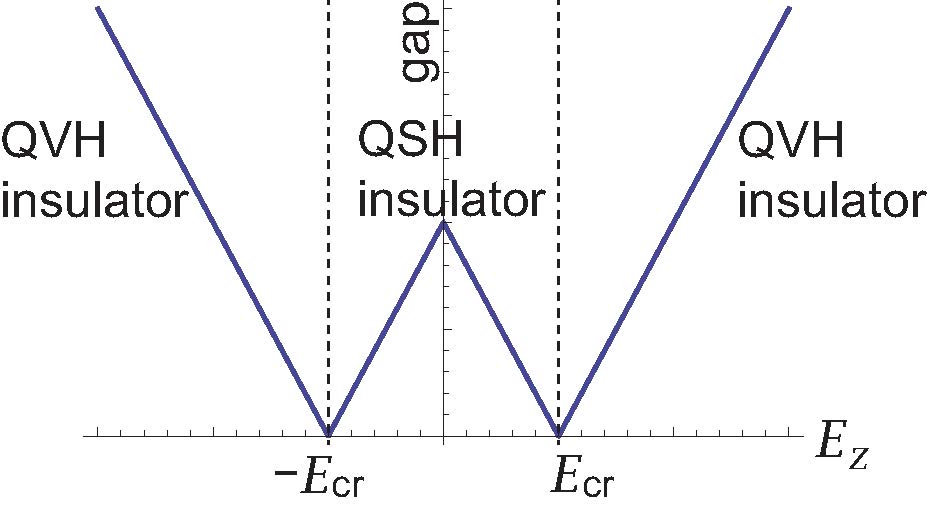}}
\caption{Electrically tunable band gap and topological phase transition of
silicene. Silicene is a QSH insulator without electric field. By applying
electric field, the band gap reduces and closes at the critical electric
field $\pm E_{\text{cr}}$. Above the critical electric field, silicene
becomes a QVH insulator. }
\label{FigGap}
\end{figure}

It is important that the band gap is tunable by controlling external
electric field $E_{z}$. The gap is open when $E_{z}=0$. As $|E_{z}|$
increases, the gap become narrower [Fig.\ref{FigGap}], and it closes at $%
E_{z}=\eta sE_{\text{cr}}$, where silicene is semimetallic just as in
graphene. As $|E_{z}|$ increases further, the gap opens again.

\subsection{Generalized Dirac mass terms}

There are actually other ways to control the band gap by introducing other
interactions to silicene. Since each Dirac cone is indexed by two parameters 
$\eta =\pm $ and $s=\pm $, the most general Dirac mass must have the
following expression,%
\begin{equation}
\Delta _{s}^{\eta }=\eta s\lambda _{\text{SO}}-\lambda _{V}+\eta \lambda
_{H}+s\lambda _{SX},  \label{DiracMass}
\end{equation}
so that it has four independent parameters, $\lambda _{\text{SO}}$, $\lambda
_{V}$, $\lambda _{H}$ and $\lambda _{SX}$. We have already discussed the
first two terms representing the SO interaction and the sublattice staggered
potential with $\lambda _{V}=\ell E_{z}$. We may write down the
tight-binding terms that yield the fourth and fifth terms\cite{2Ferro}, 
\begin{equation}
i\frac{\lambda _{H}}{3\sqrt{3}}\sum_{\left\langle \!\left\langle
i,j\right\rangle \!\right\rangle s}\nu _{ij}c_{is}^{\dagger }c_{js},\quad
\lambda _{SX}\sum_{is}s\mu _{i}c_{is}^{\dagger }c_{is}.  \label{ExtraTerms}
\end{equation}
The fourth term describes the Haldane interaction induced by the
photo-irradiation, where $\lambda _{\Omega }=\hbar v_{\text{F}}^{2}\mathcal{A}^{2}\Omega ^{-1}$ 
with $\Omega $ the frequency and $\mathcal{A}$ the
dimensionless intensity\cite{Oka,Kitagawa,EzawaPhoto}. The fifth term
describes the antiferromagnetic exchange magnetization\cite{2Ferro}.

Here we note that there are a variety of 2D materials whose low-energy
physics is described by the Dirac Hamiltonian (\ref{DiracSilic}) with the
Dirac mass (\ref{DiracMass}). We call them general honeycomb systems.
Examples are monolayer antiferromagnetic manganese chalcogenophosphates (MnPX%
$_{3}$, X = S, Se)\cite{MnPX} and perovskite G-type antiferromagnetic
insulators grown along [111] direction\cite{Hu}.

In what follows we analyze the Dirac Hamiltonian (\ref{DiracSilic}) with the
Dirac mass (\ref{DiracMass}). It can be positive, negative or zero. The band
gap is given by $2|\Delta _{s}^{\eta }|$.

\section{Topological phase transition}

\subsection{Chern numbers}

\begin{figure}[t]
\centerline{\includegraphics[width=0.45\textwidth]{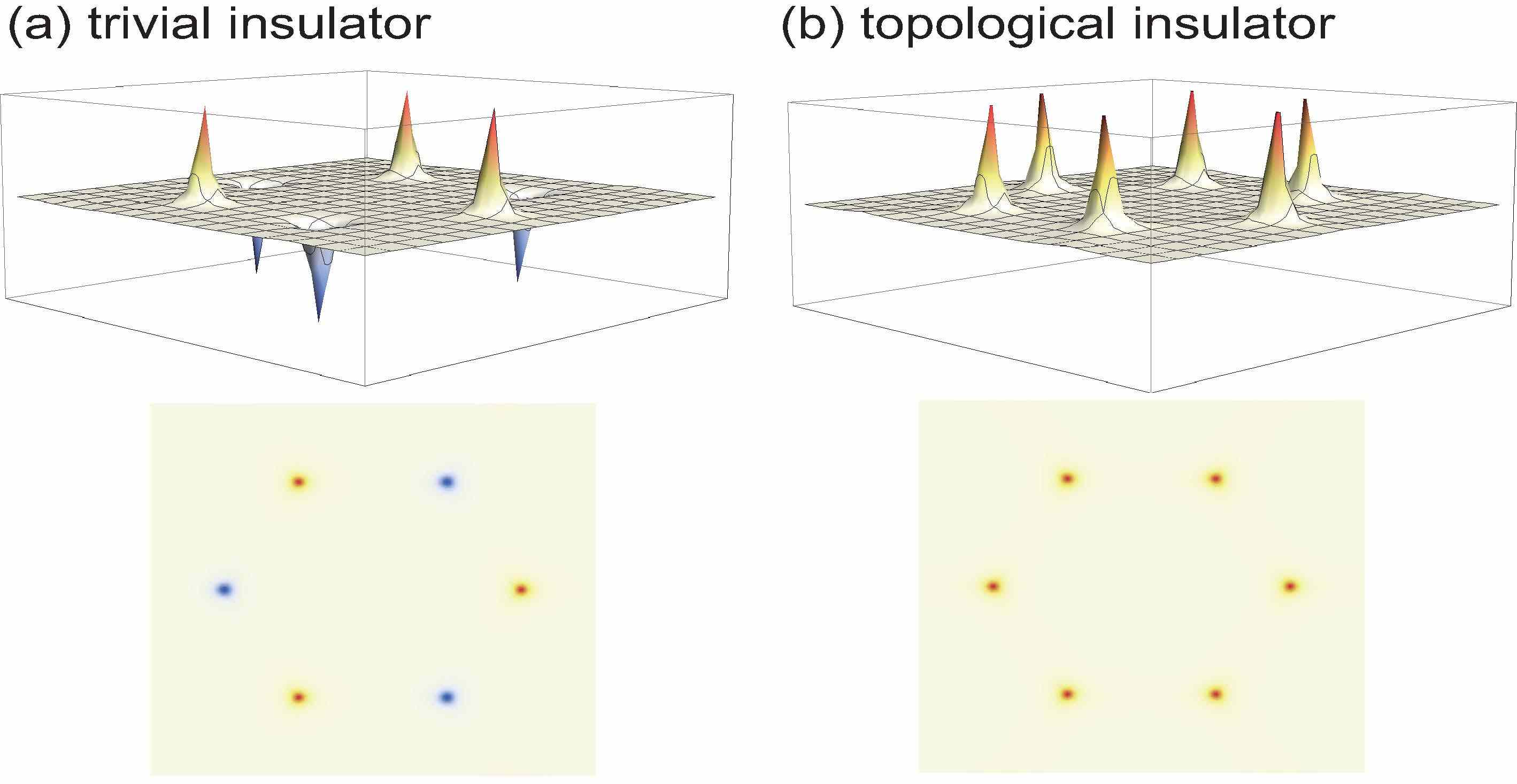}}
\caption{Berry curvature in (a) the trivial insulator and (b) the
topological insulator. It is strictly localized in the vicinity of the $K$
and $K^{\prime }$ points. The Chern number, the integration of the Berry
curvature over the Brilloin zone, is zero for a trivial insulator, while it
is non-zero for a topological insulator.}
\label{FigBerry}
\end{figure}

For any insulating state $\left\vert \psi \left( \mathbf{k}\right)
\right\rangle $ we may define a "gauge potential" in the momentum space by%
\begin{equation}
a_{k}\left( \mathbf{k}\right) =-i\left\langle \psi \left( \mathbf{k}\right)
\right\vert \partial _{k}\left\vert \psi \left( \mathbf{k}\right)
\right\rangle ,
\end{equation}%
which is properly called the Berry connection. Then we may define the
"magnetic field" in the momentum spac, which is properly called the Berry
curvature $F\left( \mathbf{k}\right) $,%
\begin{equation}
F\left( \mathbf{k}\right) =\frac{\partial }{\partial k_{x}}a_{y}\left( 
\mathbf{k}\right) -\frac{\partial }{\partial k_{y}}a_{x}\left( \mathbf{k}%
\right) .
\end{equation}%
The Chern number is the integral of the Berry curvature $F\left( \mathbf{k}%
\right) $ over the first Brillouin zone, which is the total "magnetic flux",%
\begin{equation}
\mathcal{C}=\frac{1}{2\pi }\int d^{2}kF\left( \mathbf{k}\right) .  \label{F}
\end{equation}%
We have calculated the Berry curvature with the use of the tight-binding
Hamiltonian (\ref{HamilSilic}), which we illustrate in Fig.\ref{FigBerry}.
The Berry curvature is strictly localized at the $K$ and $K^{\prime }$
points. This feature remains unchanged even if we include the extra terms (%
\ref{ExtraTerms}). Consequently, the Dirac Hamiltonian is valid to make a
topological analysis to each valley, which is indexed by the spin $%
s=\uparrow \downarrow $ and the valley index $\eta =\pm $. Namely, it is
possible to assign the Chern number $\mathcal{C}_{s}^{\eta }$ to each valley.

When the Hamiltonian is given by (\ref{DiracSilic}), the Berry curvature is
explicitly calculated for each valley as%
\begin{equation}
F_{s}^{\eta }\left( \mathbf{k}\right) =-\eta \frac{\Delta _{s}^{\eta }}{%
2\left( \left( \hbar v_{\text{F}}k\right) ^{2}+\left( \Delta _{s}^{\eta
}\right) ^{2}\right) ^{3/2}}.
\end{equation}%
The Chern number is obtained as%
\begin{equation}
\mathcal{C}_{s_{z}}^{\eta }=-\frac{\eta }{2}\text{sgn}(\Delta _{s}^{\eta }),
\label{ChernValley}
\end{equation}%
where the Dirac mass $\Delta _{s}^{\eta }$ is given by (\ref{DiracMass}).

The Chern number is quantized as $\mathcal{C}_{s}^{\eta }=\pm \frac{1}{2}$.
It is insensitive to a deformation of the band structure provided the gap is
open. On the other hand, it changes its sign as the Dirac mass $\Delta
_{s}^{\eta }$ changes its sign. Such a quantity is a topological charge.
Hence an insulator phase is indexed by a set of four Chern numbers $\mathcal{%
C}_{s}^{\eta }$. A topological phase transition occurs when the sign of the
Dirac mass $\Delta _{s}^{\eta }$ changes.

\begin{figure}[h]
\centerline{\includegraphics[width=0.45\textwidth]{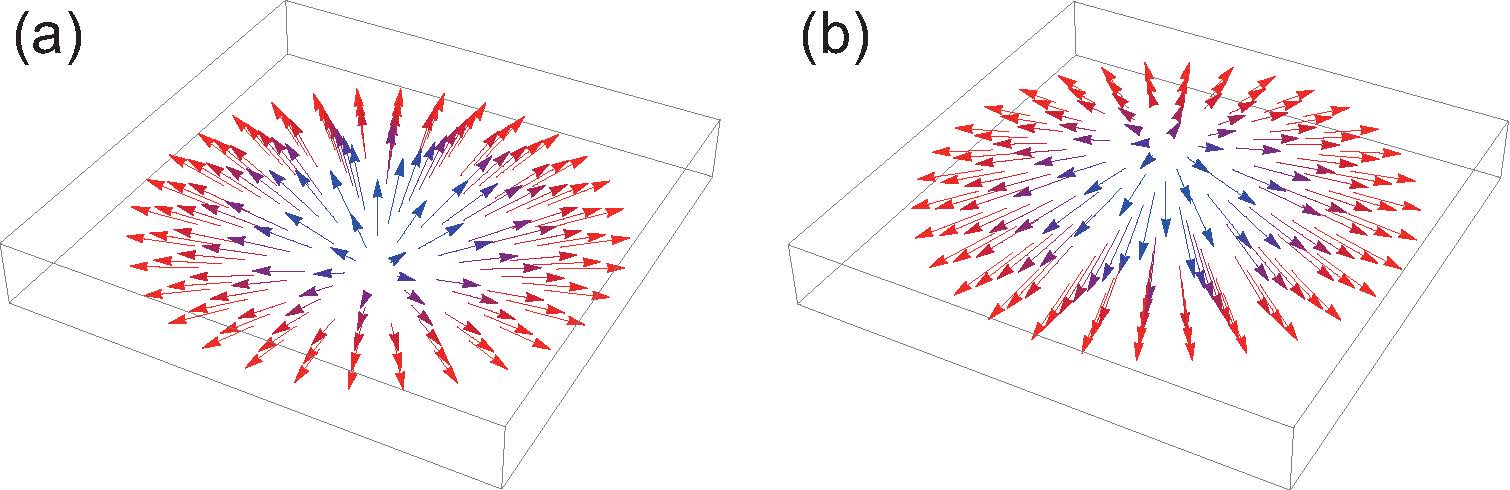}}
\caption{Illustration of a meron structure in momentum space. A meron with
the Pontryagin number (a) $1/2$, which core spin is pointing up direction,
(b) $-1/2$, which core spin is pointing down direction. }
\label{FigMeron}
\end{figure}

It is instructive to make a reinterpretation of the Chern number\cite%
{Hasan,Qi,EzawaEPJB}. When the Hamiltonian is given in terms of the $2\times
2$ Hamiltonian as in (\ref{DiracSilic}), or $H_{s}^{\eta }=\mathbf{\tau }%
\cdot \mathbf{d}$ with $d_{x}=\eta \hbar v_{\text{F}}k_{x}$, $d_{y}=\hbar v_{%
\text{F}}k_{y}$, $d_{z}=\Delta _{s}^{\eta }$, the Chern number $\mathcal{C}%
_{s}^{\eta }$ is equivalent to the Pontryagin number,%
\begin{equation}
\mathcal{C}_{s}^{\eta }=\frac{1}{4\pi }\int d^{2}k\left( \frac{\partial 
\mathbf{\hat{d}}}{\partial k_{x}}\times \frac{\partial \mathbf{\hat{d}}}{%
\partial k_{y}}\right) \cdot \mathbf{\hat{d}}.  \label{Pontryagin}
\end{equation}%
The Pontryagin number is a topological number which counts what times the
vector $\mathbf{\hat{d}}$ wraps a sphere. We use the polar coordinate of the 
$\mathbf{\hat{d}}$ vector, $\hat{d}_{x}\pm i\hat{d}_{y}=\sqrt{1-\sigma
^{2}(k)}e^{i\eta \theta }$, $\hat{d}_{z}=\sigma (k)$, and we obtain 
\begin{equation}
\mathcal{C}_{s}^{\eta }={\frac{\eta }{4\pi }}\int \!d^{2}k\;\varepsilon
_{ij}\partial _{i}\sigma \partial _{j}\theta =-{\frac{\eta }{2}}%
\int_{0}^{1}d\sigma ,
\end{equation}%
which agrees with (\ref{ChernValley}). The pseudospin texture forms a meron
structure in the momentum space. A meron is a topological structure which
has a half integer Pontryagin number as shown in Fig.\ref{FigMeron}.

\subsection{Classification of topological insulators}

We have defined four Chern numbers $\mathcal{C}_{s_{z}}^{\eta }$.
Equivalently we may define the total Chern number $\mathcal{C}$, the spin
Chern number $\mathcal{C}_{s}$\cite{Prodan,Sheng,Yang,LSheng}, the valley
Chern number\cite{Fang11,Fang13,Li} and the spin-valley Chern number\cite%
{Fang11,Kirch},%
\begin{eqnarray}
\mathcal{C} &=&\mathcal{C}_{\uparrow }^{K}+\mathcal{C}_{\uparrow
}^{K^{\prime }}+\mathcal{C}_{\downarrow }^{K}+\mathcal{C}_{\downarrow
}^{K^{\prime }},  \label{EqA1} \\
\mathcal{C}_{s} &=&\frac{1}{2}(\mathcal{C}_{\uparrow }^{K}+\mathcal{C}%
_{\uparrow }^{K^{\prime }}-\mathcal{C}_{\downarrow }^{K}-\mathcal{C}%
_{\downarrow }^{K^{\prime }}),  \label{EqA2} \\
\mathcal{C}_{v} &=&\mathcal{C}_{\uparrow }^{K}-\mathcal{C}_{\uparrow
}^{K^{\prime }}+\mathcal{C}_{\downarrow }^{K}-\mathcal{C}_{\downarrow
}^{K^{\prime }},  \label{EqA3} \\
\mathcal{C}_{sv} &=&\frac{1}{2}(\mathcal{C}_{\uparrow }^{K}-\mathcal{C}%
_{\uparrow }^{K^{\prime }}-\mathcal{C}_{\downarrow }^{K}+\mathcal{C}%
_{\downarrow }^{K^{\prime }}).  \label{EqA4}
\end{eqnarray}%
We make an important comment. The valley Chern number and the spin-valley
Chern number are well defined only in the Dirac theory. Namely they are ill
defined in the tight-binding model. Hence, we may call $\mathcal{C}$ and $%
\mathcal{C}_{s}$ the genuine Chern numbers.

Possible sets of genuine Chern numbers ($\mathcal{C},\mathcal{C}_{s}$) are $%
(0,0),(2,0),(0,1),(1,\frac{1}{2})$ up to the sign $\pm $. They are the
trivial, quantum anomalous Hall (QAH), quantum spin Hall (QSH),
spin-polarized quantum anomalous Hall (SQAH) insulators, respectively. Note
that there are two-types of trivial band insulators, which are quantum
valley Hall (QVH) insulator\cite{Tse,Qiao}, and quantum spin-valley Hall
(QSVH) insulator with antiferromagnetic (AF) order\cite{2Ferro}.

\begin{table}[tbp]
\begin{center}
\begin{tabular}{|c|c|c|c|c|c|c|c|c|}
\hline
& $\mathcal{C}_{\uparrow }^{K}$ & $\mathcal{C}_{\uparrow }^{K^{\prime }}$ & $%
\mathcal{C}_{\downarrow }^{K}$ & $\mathcal{C}_{\downarrow }^{K^{\prime }}$ & 
$\mathcal{C}$ & $2\mathcal{C}_{s}$ & $\mathcal{C}_{v}$ & $2\mathcal{C}_{sv}$
\\ \hline
QAH & $1/2$ & $1/2$ & $1/2$ & $1/2$ & $2$ & $0$ & $0$ & $0$ \\ \hline
SQAH & $1/2$ & $1/2$ & $1/2$ & $-1/2$ & $1$ & $1$ & $1$ & $-1$ \\ \hline
SQAH & $1/2$ & $1/2$ & $-1/2$ & $1/2$ & $1$ & $-1$ & $1$ & $1$ \\ \hline
QVH & $1/2$ & $1/2$ & $-1/2$ & $-1/2$ & $0$ & $0$ & $2$ & $0$ \\ \hline
SQAH & $1/2$ & $-1/2$ & $1/2$ & $1/2$ & $1$ & $1$ & $-1$ & $1$ \\ \hline
QSH & $1/2$ & $-1/2$ & $1/2$ & $-1/2$ & $0$ & $2$ & $0$ & $0$ \\ \hline
QSVH & $1/2$ & $-1/2$ & $-1/2$ & $1/2$ & $0$ & $0$ & $0$ & $2$ \\ \hline
SQAH & $1/2$ & $-1/2$ & $-1/2$ & $-1/2$ & $-1$ & $1$ & $1$ & $1$ \\ \hline
SQAH & $-1/2$ & $1/2$ & $1/2$ & $1/2$ & $1$ & $-1$ & $-1$ & $-1$ \\ \hline
QSVH & $-1/2$ & $1/2$ & $1/2$ & $-1/2$ & $0$ & $0$ & $0$ & $-2$ \\ \hline
QSH & $-1/2$ & $1/2$ & $-1/2$ & $1/2$ & $0$ & $-2$ & $0$ & $0$ \\ \hline
SQAH & $-1/2$ & $1/2$ & $-1/2$ & $-1/2$ & $-1$ & $-1$ & $1$ & $-1$ \\ \hline
QVH & $-1/2$ & $-1/2$ & $1/2$ & $1/2$ & $0$ & $0$ & $-2$ & $0$ \\ \hline
SQAH & $-1/2$ & $-1/2$ & $1/2$ & $-1/2$ & $-1$ & $1$ & $-1$ & $-1$ \\ \hline
SQAH & $-1/2$ & $-1/2$ & $-1/2$ & $1/2$ & $-1$ & $-1$ & $-1$ & $1$ \\ \hline
QAH & $-1/2$ & $-1/2$ & $-1/2$ & $-1/2$ & $-2$ & $0$ & $0$ & $0$ \\ \hline
\end{tabular}%
\end{center}
\caption{Corresponding to the spin and valley degrees of freedom, there are $%
4$ Chern numbers $\mathcal{C}_{s_{z}}^{\protect\eta }$, each of which takes $%
\pm \frac{1}{2}$, Equivalently they are given by the Chern, spin Chern,
valley Chern and spin-valley Chern numbers $\mathcal{C}$, $\mathcal{C}_{s}$, 
$\mathcal{C}_{v}$ and $\mathcal{C}_{sv}$. They are independently controlled
by the four parameters $\protect\lambda _{\text{SO}}$, $\protect\lambda _{V}$%
, $\protect\lambda _{\Omega }$ and $\protect\lambda _{SX}$. Hence there are $%
16$ states indexed by them. The genuin topological numbers are only $%
\mathcal{C}$ and $\mathcal{C}_{s}$.}
\label{tableA}
\end{table}

We comment on the relation between the $\mathbb{Z}_{2}$ index and the spin
Chern number. The spin Chern number $\mathcal{C}_{s}$ is identical to the $%
\mathbb{Z}_{2}$ index by modulo $2$ when there exists the time-reversal
symmetry\cite{Prodan}. The spin Chern number is well defined even when there
is no time-reversal symmetry, while the $\mathbb{Z}_{2}$ index is well
defined even when $s_{z}$ is not a good quantum number.

\section{\textbf{Topological edge}}

\subsection{Bulk-edge correspondence}

The most convenient way to determine if the system is topological or trivial
is to employ the bulk-edge correspondence. When there are two topological
distinct phases, a topological phase transition must occur between them. It
is generally accepted that the band gap must close at the topological phase
transition point since the topological number cannot change its quantized
value without gap closing. Note that the topological number is only defined
in the gapped system and remains unchanged for any adiabatic process.

To reveal the emergence of gapless modes at a phase transition point, it is
convenient to analyze the energy spectrum of a nanoribbon in a topological
phase, because the boundary of the nanoribbon separates a topological state
and the vacuum whose topological numbers are zero. Indeed, we have pointed
out the emergence of gapless edge modes in silicene: See Fig.\ref{FigGraSili}%
(b3). We may call it a topological edge when it separates two topologically
distinctive states.

\subsection{Helical edges and Chiral edges}

\begin{figure}[t]
\centerline{\includegraphics[width=0.5\textwidth]{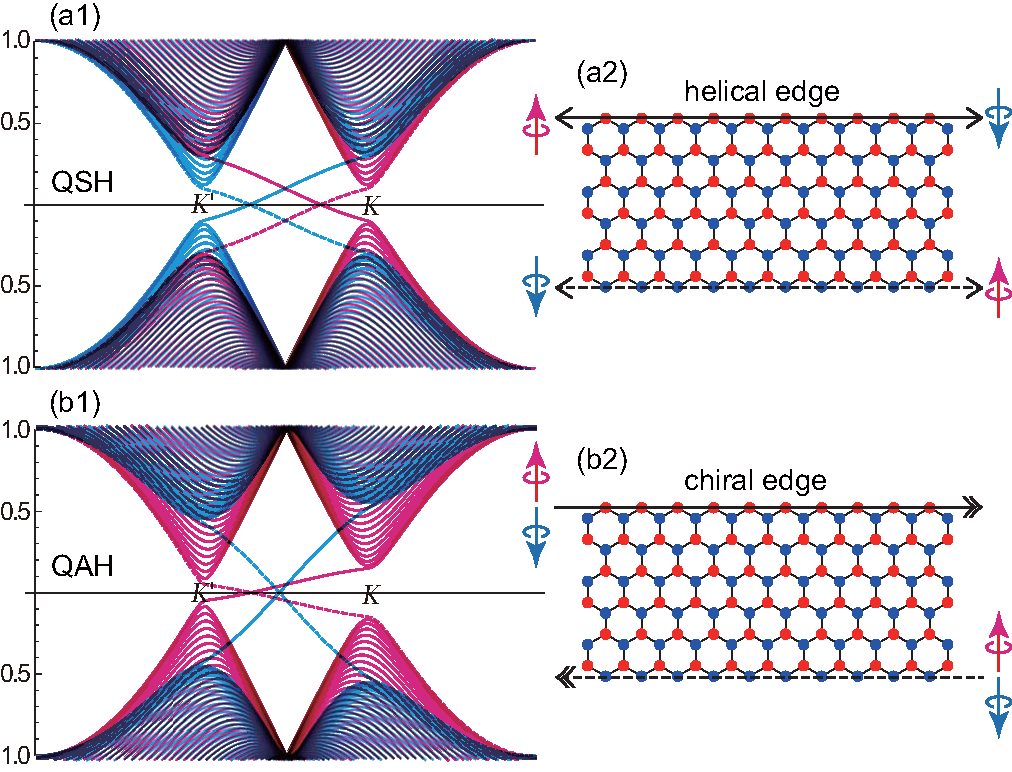}}
\caption{(a1) Nanoribbon in QSH phase. There are helical edges. We have
applied $E_{z}$ to differentiate the upper ($A$ sites) and lower ($B$ sites)
edges, which results in the splitting of the helical edges. (a2)
Illustration of helical edge states. Up and down spins flow into the
opposite directions. (b1) Nanoribbon in QAH phase. There are chiral edges.
(b2) Illustration of chiral edge states. Up and down spins flow into the
same directions. }
\label{FigSiliRibbon}
\end{figure}

We analyze the edge modes of silicene in detail when silicene is in the QSH
phase. The quantum numbers of the edge modes are not clear in Fig.\ref%
{FigGraSili}(b3) since they are degenerate. We may resolve the degeneracy by
applying week electric field $E_{z}$ to a nanoribbon with zigzag edges as in
Fig.\ref{FigSiliRibbon}(a). It is an intriguing feature of a zigzag
nanoribbon that one of the edges is composed of $A$ sites while the other of 
$B$ sites. Hence the edge modes occurring in the $A$ sites have higher
energy than those in the $B$ sites. This property enable us to identify the
edge modes occurring in the upper or lower edges. Second, the spin
degeneracy is resolved due to the SO interaction. We have shown
up(down)-spin band in magenta(cyan) in Fig.\ref{FigSiliRibbon}(a1). Third,
the velocity of electrons within the edge mode is identified by the slope of
the band gap. In this way we know how the edge current flows in Fig.\ref%
{FigSiliRibbon}(a2). A prominent feature is the up and down spins flow into
the opposite directions along each edge, implying that the edge current is a
pure spin current. Such an edge is called a helical edge.

We may also analyze a silicene nanoribbon in the QAH phase. The band
structure is given by Fig.\ref{FigSiliRibbon}(b1), where we have applied
week electric field $E_{z}$ to resolve the degeneracy. The spin and the
current direction are shown in both edges in Fig.\ref{FigSiliRibbon}(b2).
The edge current does not convey spins. Such an edge is called a chiral edge.

\subsection{Inner edges}

When a nanoribbon has only the valley Chern number and the spin-valley Chern
number, no edge modes emerge because these numbers are not defined in the
vacuum. Nevertheless these numbers are also topological numbers within the
bulk.

We may consider a junction separating two different topological phases in a
single honeycomb system\cite{EzawaNJP}. We call such a junction an inner
edge. In contrast we may call a real edge of a nanoribbon an outer edge.
There is a crucial difference between gapless edge modes appearing along an
inner edge and an outer edge. Any gapped state is indexed by a set of four
topological numbers $(\mathcal{C},\mathcal{C}_{s},\mathcal{C}_{v},\mathcal{C}%
_{sv})$. Consequently, an inner edge state carries a gapless edge mode
indexed by the difference $(\Delta \mathcal{C},\Delta \mathcal{C}_{s},\Delta 
\mathcal{C}_{v},\Delta \mathcal{C}_{sv})$ between the two adjacent gapped
states. More precisely, we set $\Delta \mathcal{C=C}^{L}-\mathcal{C}^{R}$
and so on, when the topological insulator with $(\mathcal{C}^{L},\mathcal{C}%
_{s}^{L},\mathcal{C}_{v}^{L},\mathcal{C}_{sv}^{L})$ is on the left-hand side
of the one with $(\mathcal{C}^{R},\mathcal{C}_{s}^{R},\mathcal{C}_{v}^{R},%
\mathcal{C}_{sv}^{R})$. On the other hand, an outer edge state can carry a
gapless edge mode only indexed by $(\mathcal{C},\mathcal{C}_{s})$ of the
gapped state because the valley Chern numbers are ill defined in the vacuum.

\begin{figure}[t]
\centerline{\includegraphics[width=0.5\textwidth]{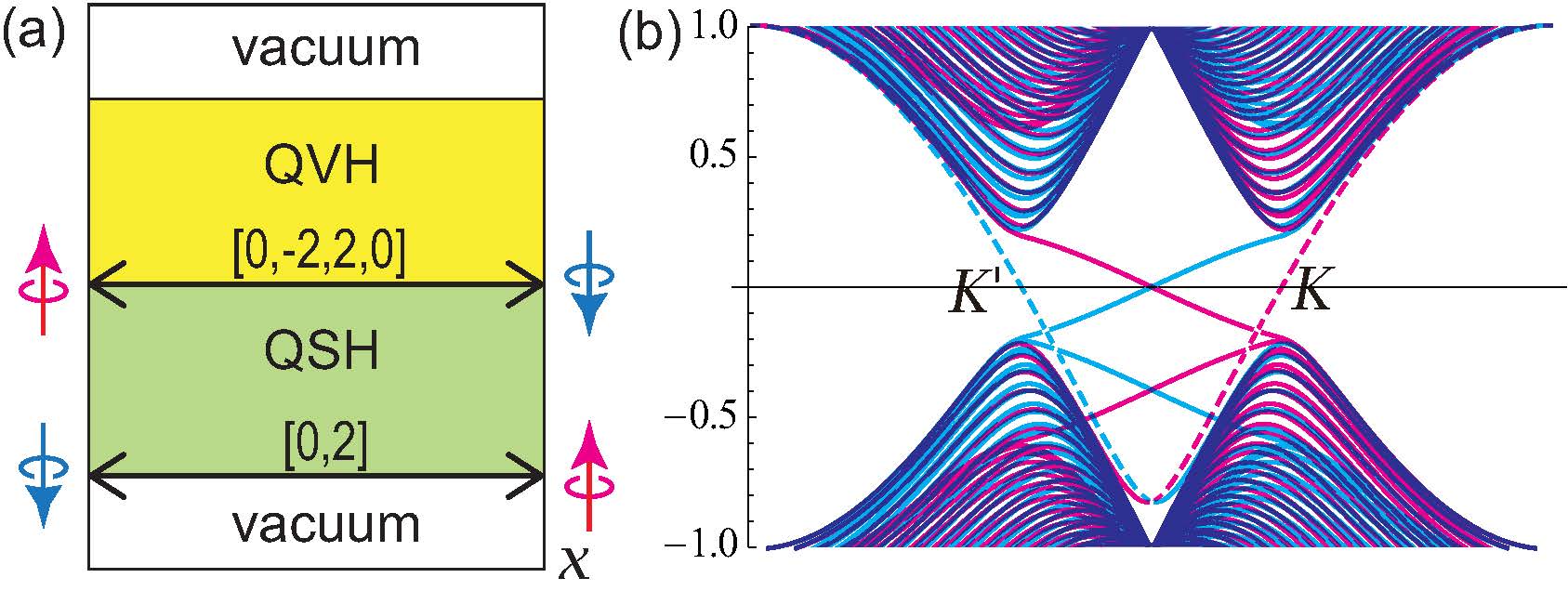}}
\caption{(a) Silicene nanoribbon placed parallel to the $x$ axis. When
external field $E_{z}$ is applied along the $y$ axis, the region with $%
E_{z}>E_{\text{cr}}$ becomes a QVH insulator. The inner edge between the QSH
and QVH parts is helical. The outer edge below the QSH part is also helical.
No gapless edge states appear along the outer edge above the QVH part. (b)
The band structure of a silicene nanoribbon. Four gapless edge modes are
found, which are assigned as in (a). }
\label{FigInnerEdge}
\end{figure}

We illustrate a nanoribbon which contains the QSH and QVH phases in Fig.\ref%
{FigInnerEdge}, where there are two outer edges and one inner edge. As we
have argued, the outer edge of the QSH part is helical, while the outer edge
of the QVH part has no gapless edge modes. The nature of the inner edge is
seen by analyzing the band structure in Fig.\ref{FigInnerEdge}(b). It
contains only four nondegenerate gapless states. On one hand, two solid
lines correspond to the helical edge between the QSH part and the vacuum. On
the other hand, two dotted lines correspond to the inner edge between the
QSH and QVH parts, which is also helical. It should be noted that the inner
edge modes cross the Fermi energy at the $K$ and $K^{\prime }$ points.

\subsection{\textbf{Topological Kirchhoff law}}

\begin{figure}[t]
\centerline{\includegraphics[width=0.5\textwidth]{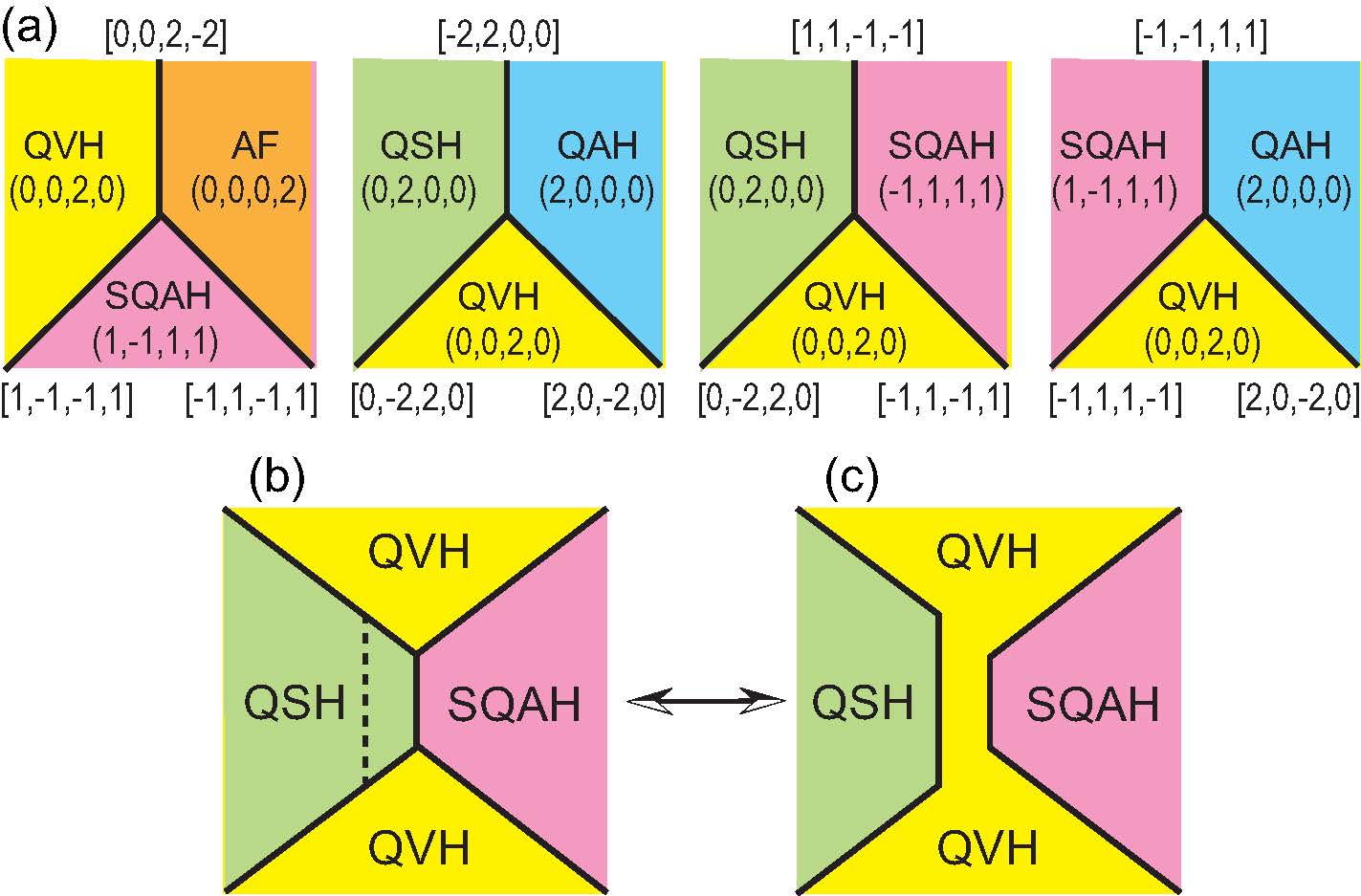}}
\caption{(a) Y junctions of topological edge currents, which are made of the
boundaries of three different topological insulators. (b) and (c) Topological
circuits made of topological edges. We apply electric field to the center
region of the QSH phase in (b). The region, to which electric field is applied, becomes a
QVH phase. As a result, the connected edge states are separated as shown in (c).}
\label{FigYJunc}
\end{figure}

We consider a configuration where three different topological insulators
meet at one point: See Fig.\ref{FigYJunc}. In this configuration there are
three edges forming a Y-junction. The condition which edges can make a
Y-junction is the conservation of these topological numbers at the junction.
This law is a reminiscence of the Kirchhoff law, which dictates the
conservation of currents at the junction of electronic circuits. We call it
the topological Kirchhoff law\cite{Kirch}.

We present an interesting interpretation of the topological Kirchhoff law.
We may regard each topological edge state as a world line of a particle
carrying the four topological charges. The Y-junction may be interpreted as
a scattering process of these particles. In this scattering process, the
topological charges conserve. We have shown that we can control the mass of
Dirac cones with the spin and valley independently in silicene. Our findings
will open a new way to topological spin-valleytronics, where the spin and
valley degrees of freedom and the topology are fully manipulated.

\section{\textbf{Gapless edge mode in Dirac theory}}

We construct the Dirac theory of a gapless inner-edge state\cite%
{EzawaNJP,Kirch} indexed by $\eta $ and $s$. It emerges along a curve where
the Dirac mass vanishes, $\Delta _{s}^{\eta }\left( x,y\right) =0$.

\subsection{Rectangular geometry}

We take the edge along the $x$ axis. Zero modes emerge along the line
determined by $\Delta _{s}^{\eta }\left( y\right) =0$. We may set $k_{x}=$%
constant due to the translational invariance along the $x$ axis. The
equation of motion $H_{K}\psi =0$ reads 
\begin{equation}
\left( -i\hbar v_{\text{F}}\tau _{y}\partial _{y}+\Delta _{s}^{\eta }\tau
_{z}\right) \psi =0.
\end{equation}%
We set $\psi _{B}=i\zeta \psi _{A}$ with $\zeta =\pm 1$, and seek the
zero-energy solution. Here, $\psi _{A}$ is a two-component amplitude with up
and down spins, $\psi _{A}=(\psi _{A}^{\uparrow },\psi _{A}^{\downarrow })$.
Furthermore, setting $\psi _{A}\left( x,y\right) =e^{ik_{x}x}\phi _{A}\left(
y\right) $, we obtain $H_{\eta }\psi _{A}\left( x,y\right) =E_{\eta \zeta
}\psi _{A}\left( x,y\right) $ together with $E_{\eta \zeta }=\eta \zeta
\hbar v_{\text{F}}k_{x}$, and 
\begin{equation}
\left( \xi \hbar v_{\text{F}}\partial _{y}-\Delta _{s}^{\eta }\left(
y\right) \right) \phi _{A}(y)=0.
\end{equation}%
By solving this, the zero-energy solution is given by 
\begin{equation}
\psi _{A}\left( x,y\right) =Ce^{ik_{x}x}\exp \left[ \frac{\zeta }{\hbar v_{%
\text{F}}}\int^{y}\Delta _{s}^{\eta }\left( y^{\prime }\right) dy^{\prime }%
\right] ,  \label{ZeroModeSolut}
\end{equation}%
and $\psi _{B}\left( x,y\right) =i\zeta \psi _{A}\left( x,y\right) $, where $%
C$ is the normalization constant. The sign of $\zeta =\pm 1$ is determined
so as to make the wave function finite in the limit $\left\vert y\right\vert
\rightarrow \infty $. This is a reminiscence of the Jackiw-Rebbi mode\cite%
{Jakiw} presented for the chiral mode. The difference is the presence of the
spin and valley indices in the wave function.

\subsection{Circular geometry}

We consider a cylindrical symmetric domain with the radius $r_{0}$ at the
origin of the $xy$ plane. A phase transition occurs at $r=r_{0}$, where $%
\Delta _{s}^{\eta }(r)=0$. The equation of motion $H_{K}\psi =0$ reads%
\begin{eqnarray}
\Delta _{s}^{\eta }(r)\psi _{s,A}^{\eta }+\hbar v_{\text{F}}e^{i\eta \theta
}\left( i\partial _{r}-\frac{1}{r}\partial _{\theta }\right) \psi
_{s,B}^{\eta } &=&0,  \notag \\
\hbar v_{\text{F}}e^{-i\eta \theta }\left( i\partial _{r}+\frac{1}{r}%
\partial _{\theta }\right) \psi _{s,A}^{\eta }-\Delta _{s}^{\eta }(r)\psi
_{s,B}^{\eta } &=&0.  \label{EqA}
\end{eqnarray}%
We can solve this for zero-energy states as%
\begin{equation}
\psi _{s,A}^{\eta }(r,\theta )=\frac{C}{\sqrt{r}}e^{i\eta \theta /2}\exp %
\left[ \frac{\xi }{\hbar v_{\text{F}}}\int_{0}^{r}\Delta _{s}^{\eta
}(r)dr^{\prime }\right] ,  \label{Circle}
\end{equation}%
and $\psi _{B}\left( r,\theta \right) =i\zeta \psi _{A}\left( r,\theta
\right) $, where $C$ is the normalization constant and $\zeta =\pm 1$. The
sign of $\zeta $ is determined so as to make the wave function finite in the
limit $r\rightarrow \infty $.

\subsection{Interface induced by electric field}

We apply an inhomogeneous electric field, 
\begin{equation}
\Delta _{s}^{\eta }(r)=\lambda _{\text{SO}}-\lambda _{V}\left( r\right) 
\end{equation}%
with $\lambda _{V}\left( r\right) =\ell E_{z}\left( r\right) >\lambda _{%
\text{SO}}$ for $r<r_{0}$. The region $r<r_{0}$ is a trivial insulator,
while the region $r>r_{0}$ is a quantum spin-Hall insulator. The zero-energy
helical edge current flows along the circle $r=r_{0}$.

\section{Topological quantum field-effect transistor}

\begin{figure}[t]
\centerline{\includegraphics[width=0.5\textwidth]{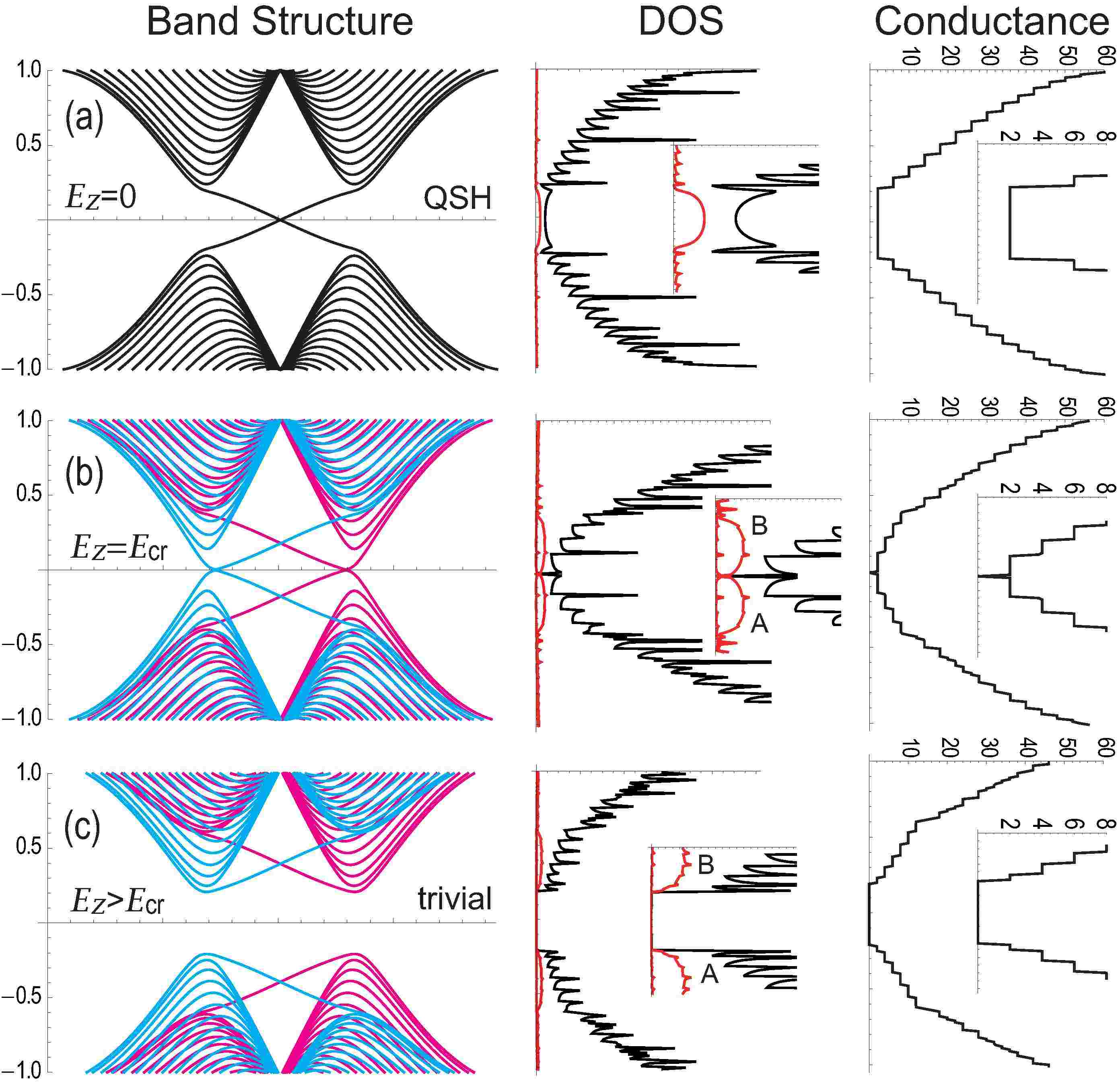}}
\caption{(Color online) Band structure, DOS and conductance of zigzag
silicene nanoribbons for (a) the QSH insulator phase, (b) the metalic phase
at the phase transition point, and (c) the trivial insulator phase. These
phases are obtained by applying electric field $E_{z}$. The phase transition
occurs at $E_{z}=E_{\text{cr}}$. The number of bands is $2W+2$ in the
nanoribbon with width $W$. Here, the width is taken to be $W=31$, and only a
part of bands are shown. The band gap is degenerate (nongenerate) with
respect to the up (red) and down (blue) spins at $E_{z}=0$ ($E_{z}>0$). Van
Hove singularities emerge in the DOS at the points where the band dispersion
is flat. The site-resolved DOS of the up-spin state at the outmost $A$ and $%
B $ sites of a nanoribbon are shown by red curves in the insets. There are
finite DOS for the zero-energy edge states in the QSH insulator. The
conductance is quantized by unit of $e^{2}/h$. }
\label{FigESili}
\end{figure}

We next calculate the conductance of a nanoribbon by using the Landauer
formalism\cite{Datta,EzawaAPL}. The conductance is quantized in silicene
nanoribbons. Indeed, one channel has a quantized conductance $e^{2}/h$.
Accordingly, the conductance is obtained by counting the number of bands. We
show the conductance in Fig.\ref{FigESili}. When electric field is not
applied, there are helical edge states, which contribute to the conductance $%
2e^{2}/h$ since up and down spin channels contribute to the conductance.
When the electric field $E_{z}$ exceeds the critical value $E_{\text{cr}}$,
the edge states disappear since the nanoribbon becomes a trivial insulator,
which results in zero conductance. This means the system acts as a
transistor where the "on" state can be switched to the "off" state by
applying electric field. This transistor is "quantum" since the conductance
is quantized, which is highly contrasted with the ordinal transistor, where
the conductance is not quantized. Furthermore the conductance is
topologically protected because the zero-energy edge state is topologically
protected. Namely the conductance is robust against impurities due to its
topological stability. Consequently we may call it a field-effect
topological quantum transistor\cite{EzawaAPL}. This is the most
energy-saving device since it utilizes the minimum conductance.

We have calculated the density of states (DOS) $\rho (E)$ and the
conductance $\sigma (E)$ of a nanoribbon as functions of the Fermi energy $E$%
, which is controlled by doping. We give the results at electric field $%
E_{z}=0$, $E_{\text{cr}}$ and $2E_{\text{cr}}$ in Fig.\ref{FigESili}. A van
Hove singularity occurs in the DOS at the point where the band dispersion is
flat. As $E$ increases beyond the point, the Fermi level crosses a new band.
A new channel opens and contributes to the conductance by $e^{2}/h$ for each
spin and valley. It is clearly observed that the edge channel connects the
tips of the Dirac cones with the same spin at the $K$ and $K^{\prime }$
points. 

We have also plotted the site-resolved DOS $\rho _{i}(E)$ of the up-spin
states at the outmost $A$ and $B$ sites of a nanoribbon by red curves in the
insets [Fig.\ref{FigESili}]. They represent degenerate zero-energy states at 
$E_{z}=0$. As we have explained in Fig.\ref{FigSiliRibbon}, the energy of
the $A$ and $B$ sites become different for $E_{z}\neq 0$. It results in the
downward (upward) shift of $\rho _{A(B)}(E)$ along the edge as $E_{z}$
increases. They are separated completely, and zero-energy states disappear
for $E_{z}>E_{\text{cr}}$.

\section{Quantum anomalous Hall effects}

Silicene has an additional interaction term, which we have so far neglected.
It is the Rashba SO interaction given by%
\begin{equation}
-i\frac{2}{3}\lambda _{\text{R}}\sum_{\left\langle \!\left\langle
i,j\right\rangle \!\right\rangle \alpha \beta }\mu _{i}c_{i\alpha }^{\dagger
}\left( \mathbf{\sigma }\times \hat{\mathbf{d}}_{ij}\right) _{\alpha \beta
}^{z}c_{j\beta }  \label{Rashba}
\end{equation}%
in the tight-binding model. Here, $\lambda _{\text{R}}$ represents the
Rashba SO coupling strength associated with next-nearest neighbor hopping,
where $\mu _{i}=\pm 1$ for the A (B) site, and $\hat{\mathbf{d}}_{ij}=%
\mathbf{d}_{ij}/\left\vert \mathbf{d}_{ij}\right\vert $ with the vector $%
\mathbf{d}_{ij}$ connecting two sites $i$ and $j$ in the same sublattice\cite%
{LiuPRB}. The Dirac theory of the Rashba term is given by%
\begin{equation}
\eta \tau _{z}a\lambda _{\text{R}}\left( k_{y}\sigma _{x}-k_{x}\sigma
_{y}\right) .
\end{equation}%
The Rashba term vanishes at the $K$ and $K^{\prime }$ points. Hence the
Rashba interaction is negligible as far as the low-energy physics near the
Dirac points is concerned.

\begin{figure}[t]
\centerline{\includegraphics[width=0.5\textwidth]{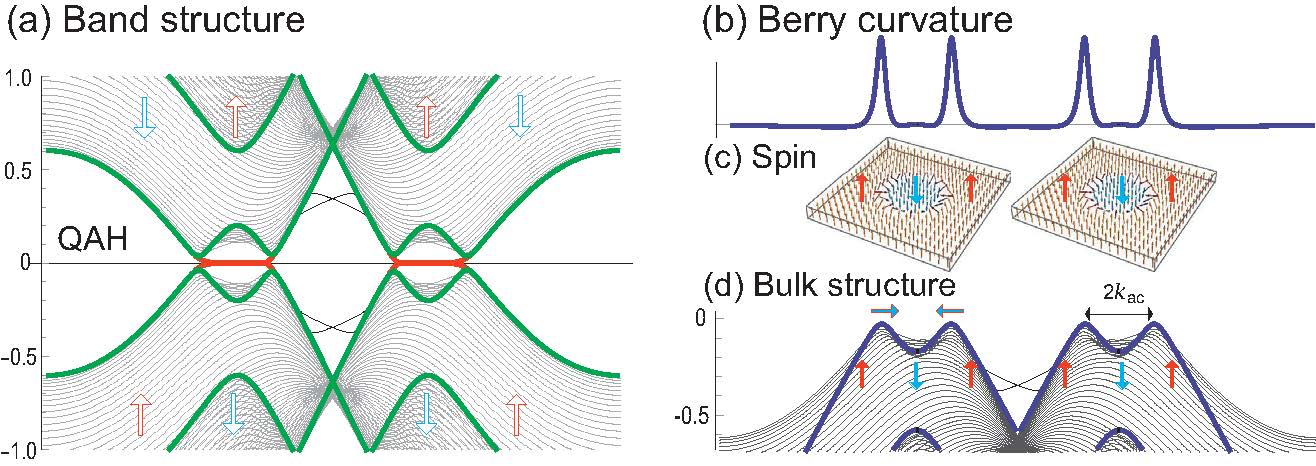}}
\caption{(a) The energy spectrum of a QAH state. Gray curves are for a
nanoribbon. Heavy green curves represent the energy spectrum of the bulk,
calculated independently. A gap opens in the bulk spectrum, where almost
flat gapless modes appear at the edges of a nanoribbon. A red (blue) arrow
indicates the spin direction away from the Fermi level. (b) (a) Berry
curvature, (b) spin, and (c) band structure of a QAH state calculated based
the Dirac theory. Gray curves represent the energy spectrum of a nanoribbon.
Spins rotate by the Rashba interaction near the Fermi level, generating a
skyrmion spin texture in the momentum space. It generates a nontrivial Berry
curvature along the anticrossing circle. The integration of the Berry
curvature gives the Chern number $C=2$, since there are two skyrmions each
of which yields $C=1$. }
\label{FigQAH}
\end{figure}

There exists an exceptional case in which we cannot neglect the Rashba
interaction. We have previously studied the interaction term (\ref{DiracMass}%
) affecting the Dirac mass. There is another type of interactions, which do
not contribute to the Dirac mass but shift the bands into the opposite
directions between up and down spins. The most important one is given by the
exchange interaction,%
\begin{equation}
M\sum_{is}sc_{is}^{\dagger }c_{is}\qquad \text{or}\qquad M\sigma _{z}
\end{equation}%
in the tight-binding model or in the Dirac theory. We show the band
structure of a nanoribbon with the exchange interaction in Fig.\ref{FigQAH}%
(a). We see that the two Dirac cones with the opposite spins approach as $|M|
$ increases and touch each other at $|M|=\lambda _{\text{SO}}$. Then, we
expect naively the level crossing to occur for $\left\vert M\right\vert
>\lambda _{\text{SO}}$, that is, the two Dirac cones to penetrate into each
other. Actually, the level crossing turns into the level anticrossing due to
the spin mixing caused by the Rashba interaction $\lambda _{\text{R}}$. The
new phase is the QAH insulator\cite{EzawaQAH} with the Chern number $2$. We
show the Berry curvature in Fig.\ref{FigQAH}(b), where it takes a large
value where the bands almost touch the Fermi energy away from the $K$ and $%
K^{\prime }$ points. The spin direction is inverted and the spin forms a
skyrmion texture, which results in the Chern number $1$. The total Chern
number is $2$ since there are two skyrmions at the $K$ and $K^{\prime }$
points. 

\section{Symmetry protected topological charge}

The Chern number $\mathcal{C}$ is always quantized. However, the other ones $%
\mathcal{C}_{i}$ are not so when the relevant symmetry is broken. Indeed,
the spin $s_{z}$ symmetry is broken by the Rashba interaction. We here
analyze this problem.

There exists an alternative expression for the Chern number in terms of the
Green function\cite{Volovik}. It reads%
\begin{eqnarray}
\mathcal{C} &=&\frac{\varepsilon _{\alpha \beta \gamma }}{6}\int \frac{d^{2}k%
}{\left( 2\pi \right) ^{2}}\int_{-\infty }^{\infty }d\omega \,  \notag \\
&&\times \text{Tr}[G\partial _{k_{\alpha }}G^{-1}G\partial _{k_{\beta
}}G^{-1}G\partial _{k_{\gamma }}G^{-1}],  \label{GdG}
\end{eqnarray}%
where $k_{\alpha }$, $k_{\beta }$ and $k_{\gamma }$ run through $k_{0}\equiv
i\omega $, $k_{x}$ and $k_{y}$, while $G\left( \omega ,\mathbf{k}\right) $
is the Matsubara Green's function, 
\begin{equation}
G\left( \mathbf{k}\right) =\frac{1}{i\omega -H\left( \mathbf{k}\right) },
\end{equation}%
with $i\omega $ referring to the Matsubara frequency ($\omega $: real). This
formula has a merit that it can be used even in the presence of interactions%
\cite{Wang,Wang12,Gurarie}.

We propose to define the following quantities\cite{EzawaSPT},%
\begin{eqnarray}
\mathcal{\tilde{C}}_{i} &=&\frac{\varepsilon _{\alpha \beta \gamma }}{6}\int 
\frac{d^{2}k}{\left( 2\pi \right) ^{2}}\int_{-\infty }^{\infty }d\omega \, 
\notag \\
&&\times \text{Tr}[\chi _{i}G\partial _{k_{\alpha }}G^{-1}G\partial
_{k_{\beta }}G^{-1}G\partial _{k_{\gamma }}G^{-1}],  \label{SpinChern}
\end{eqnarray}%
for $i=s,v,sv$ together with $\chi _{s}=\sigma _{z}$, $\chi _{v}=\eta _{z}$, 
$\chi _{sv}=\sigma _{z}\eta _{z}$. Provided the spin $\sigma _{z}$ and the
pseudospin $\eta _{z}$ are good quantum numbers, namely, $\left[ H,\sigma
_{z}\right] =\left[ H,\eta _{z}\right] =0$, we are able to prove that these
are identical to (\ref{EqA2}), (\ref{EqA3}) and (\ref{EqA4}) by following
the method\cite{ZhangB}. We may use the formula (\ref{SpinChern}) even if
symmetries are broken. They are no longer quantized when \ the symmetries
are broken. Accordingly, we call them symmetry protected topological charges.

In the presence of the Rashba interaction, the spin Chern number is
explicitly calculated as%
\begin{equation*}
\mathcal{\tilde{C}}_{s}=\frac{\mathcal{C}_{s}}{\left[ 1+(a\lambda _{\text{R}%
}/\hbar v_{\text{F}})^{2}\right] },
\end{equation*}%
where $\mathcal{C}_{s}$ is the spin Chern number without the Rashba
interaction. This yields $\mathcal{\tilde{C}}_{s}=1-5.9\times 10^{-7}$,
where we have used $v_{\text{F}}=5.5\times 10^{5}$m/s, $a=3.86$\AA\ and $%
\lambda _{\text{R2}}=0.7$meV as sample parameters of silicene. Surely $%
\mathcal{\tilde{C}}_{s}$ is not quantized, but the deviation is negligibly
small. Furthermore, we can check that the edge mode remains practically
gapless. The spin mixing can be neglected in practical purposes.

\section{\textbf{Topological superconductor}}

Topological superconductors are superconductors which have nontrivial
topological numbers\cite{Hasan,Qi}. The bulk band spectrum has a full gap
due to the superconducting gap, while the edge states appear at the sample
of edges as in the case of topological insulators. An intriguing feature is
that these edge states support Majorana fermions\cite{Alicea,Beenaker}.

\subsection{Majorana fermion}

Majorana fermion is one of the hottest topics in the condensed matter
physics. Majorana fermion operator $\gamma $ satisfies the anti-commutation
relation $\left\{ \gamma ,\gamma ^{\dagger }\right\} =1$ and the Majorana
condition $\gamma =\gamma ^{\dagger }$, indicating that a Majorana fermion
is a particle which is its own anti-particle.

In the BCS theory, superconductor is expressed by the Bogoliubov--de Gennes
(BdG) Hamiltonian,%
\begin{equation}
H=\frac{1}{2}\sum_{k}\left( c_{k}^{\dagger },c_{-k}\right) H_{\text{BdG}%
}\left( k\right) \left( 
\begin{array}{c}
c_{k} \\ 
c_{-k}^{\dagger }%
\end{array}%
\right) ,
\end{equation}%
based on the Nambu representation $\left( c_{k},c_{-k}^{\dagger }\right) $,
which is a combination of the particle and hole operators. The BdG
Hamiltonian has a particle-hole symmetry,%
\begin{equation}
\Xi H_{\text{BdG}}\Xi ^{-1}=-H_{\text{BdG}},
\end{equation}%
with respect to the particle-hole operator $\Xi $ satisfying $\Xi ^{2}=1$.
The states with the energy $E$ and $-E$ are to be identified in the Nambu
representation. The creation and annihilation operators of these two states
satisfy $\gamma _{E}^{\dagger }=\gamma _{-E}$. Consequently, we obtain $%
\gamma _{0}^{\dagger }=\gamma _{0}$ for a single zero-energy state ($E=0$)
if there is such one. As a result, a single zero-energy state with the
particle-hole symmetry is a Majorana fermion.

\subsection{Topological superconductor in honeycomb system}

A topological superconductor is obtained from a QAH insulator by attaching
an $s$-wave superconductor to it\cite{QiQAH}. Indeed, Cooper pairs are
formed between up and down spins at the same site of the honeycomb system.
The resultant BCS Hamiltonian reads $H_{\text{BCS}}=H_{K}+H_{K^{\prime }}+H_{%
\text{SC}}$ with%
\begin{align}
\hat{H}_{\text{SC}}=& \sum_{\tau =A,B}[\Delta _{\text{SC}}c_{\tau \uparrow
}^{K\dagger }\left( k\right) c_{\tau \downarrow }^{K^{\prime }\dagger
}\left( -k\right) +\Delta _{\text{SC}}c_{\tau \uparrow }^{K^{\prime }\dagger
}\left( k\right) c_{\tau \downarrow }^{K\dagger }\left( -k\right) ]  \notag
\\
& +\text{h.c.}  \label{SilicBCS}
\end{align}%
in the momentum representation, where $\Delta _{\text{SC}}$\ is the
superconducting gap. A finite gap present in a superconducting state allows
us to evaluate the Chern number of the state to determine whether it is
topological. Alternatively we may examine the emergence of gapless edge
modes by calculating the band structure of a nanoribbon with zigzag edge
geometry based on this Hamiltonian. The BCS Hamiltonian is rewritten into
the BdG Hamiltonian,%
\begin{equation}
H_{\text{BdG}}=\left( 
\begin{array}{cc}
H_{K}\left( k\right) & H_{\Delta } \\ 
H_{\Delta }^{\dagger } & -H_{K^{\prime }}^{\ast }\left( -k\right)%
\end{array}%
\right) ,  \label{BdG1}
\end{equation}%
by introducing the Nambu representation for the basis vector, i.e., $\Psi
=\left\{ \psi _{A\uparrow }^{K},\psi _{B\uparrow }^{K},\psi _{A\downarrow
}^{K},\psi _{B\downarrow }^{K},\psi _{A\uparrow }^{K^{\prime }\dagger },\psi
_{B\uparrow }^{K^{\prime }\dagger },\psi _{A\downarrow }^{K^{\prime }\dagger
},\psi _{B\downarrow }^{K^{\prime }\dagger }\right\} ^{t}$.

Diagonalizing the BdG Hamiltonian, we obtain the energy spectrum. It
consists of eight levels with the eigenvalues%
\begin{equation}
E_{\text{BdG}}^{\alpha ,\beta }\left( k\right) =\pm \sqrt{\left( \hbar v_{%
\text{F}}k\right) ^{2}+\left( E_{0}^{\alpha ,\beta }\right) ^{2}}
\label{EnergA}
\end{equation}%
with%
\begin{equation}
E_{0}^{\alpha ,\beta }=\sqrt{\left( \left( \lambda _{\text{SO}}-\alpha
\lambda _{V}\right) ^{2}+\Delta _{\text{SC}}^{2}\right) }+\beta \left(
\lambda _{\text{H}}+\alpha \lambda _{\text{SX}}\right) ,  \label{EqC}
\end{equation}%
where $\alpha $ and $\beta $ takes $\pm 1$. The gap closes ($E_{0}^{\alpha
,\beta }=0$) at%
\begin{equation}
\left( \lambda _{\text{H}}+\alpha \lambda _{\text{SX}}\right) ^{2}=\left(
\lambda _{\text{SO}}-\alpha \lambda _{V}\right) ^{2}+\Delta _{\text{SC}}^{2}.
\label{boundary}
\end{equation}%
Though the original Hamiltonian is an $8\times 8$ matrix, we may decompose
it into 4 independent $2\times 2$ Hamiltonians,%
\begin{equation}
H^{\alpha ,\beta }\left( k\right) =\left( 
\begin{array}{cc}
\beta E_{0}^{\alpha ,\beta } & \hbar v_{\text{F}}k_{-} \\ 
\hbar v_{\text{F}}k_{+} & -\beta E_{0}^{\alpha ,\beta }%
\end{array}%
\right) ,  \label{BdG}
\end{equation}%
corresponding to $\alpha ,\beta =\pm 1$. This Hamiltonian reproduces the
energy spectrum (\ref{EnergA}). We may interpret $\beta E_{0}^{\alpha ,\beta
}$ as the modified Dirac mass due to the BCS condensation.

It is straightforward to calculate the Chern number of the superconducting
honeycomb system $H_{\text{BdG}}$. It is determined by the sign of the
modified Dirac mass\cite{EzawaEPJB},%
\begin{equation}
C=\frac{1}{2}\sum_{\alpha ,\beta =\pm 1}\text{sgn}\left( \beta E_{0}^{\alpha
,\beta }\right) .  \label{Chern}
\end{equation}%
The condition of the emergence of a topological superconductivity is $C\neq
0 $. Note that it is zero when the time-reversal symmetry is present. In
order to obtain a non-zero Chern number, $\lambda _{\text{SX}}$ or $\lambda
_{\text{H}}$ must be nonzero. It should be noticed that the spin Chern
number is no longer defined due to the BCS condensation of the up-spin and
down-spin electrons. The topological phase diagram is easily constructed in
the $(\lambda _{\text{SO}},\lambda _{V},\lambda _{\text{SX}},\lambda _{\text{%
H}},\Delta _{\text{SC}})$ space. The phase boundaries are given by (\ref%
{boundary}). The Chern number is determined from (\ref{Chern}).

\begin{figure}[t]
\centerline{\includegraphics[width=0.4\textwidth]{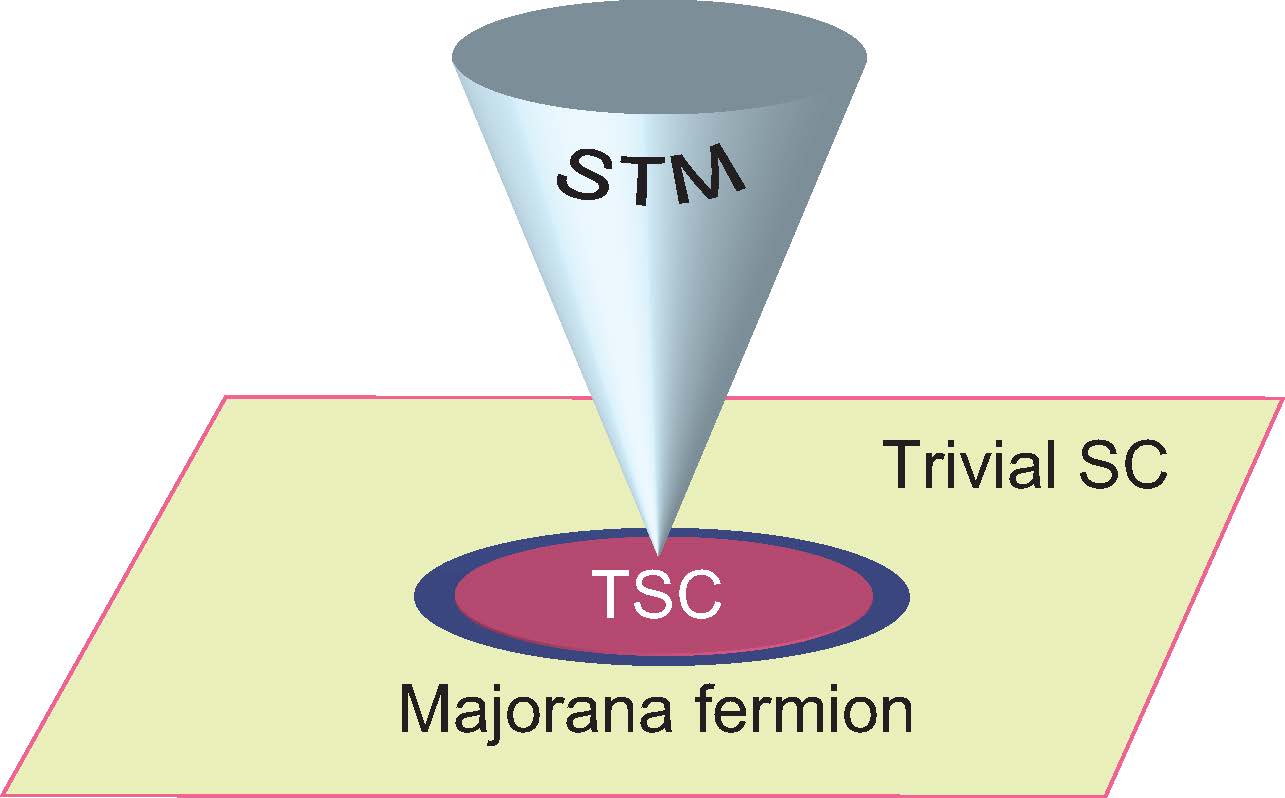}}
\caption{Illustration of a Majorana zero-energy state between two
topological phases. By applying electric field $E_{z}$ locally, we may
create a topological spot ($C=1$) in a trivial superconductor ($C=0$). There
appears a zero-energy Majorana state at the phase boundary.}
\label{pond}
\end{figure}

\subsection{Majorana fermion in honeycomb system}

We consider a disk region in a honeycomb sheet, as illustrated in Fig.\ref%
{pond}. We may tune parameters $\lambda _{\text{SO}}$, $\lambda _{V}$, $%
\lambda _{\text{SX}}$, $\lambda _{\text{H}}$ and $\Delta _{\text{SC}}$ to
become space-dependent so that the inner region has a different Chern number
from the outer region. There appears gapless edge modes at the phase
boundary, where $E_{0}^{\alpha ,\beta }(r)=0$. According to a general
theorem, as we have reviewed, a zero-energy state with the particle-hole
symmetry is always a Majorana fermion.

There are several ways to make $E_{0}^{\alpha ,\beta }(r_{c})=0$, since
there are four independent mass parameters $\lambda _{\text{SO}}$, $\lambda
_{V}$, $\lambda _{\text{SX}}$, $\lambda _{\text{H}}$ and one superconducting
gap $\Delta _{\text{SC}}$. A simple way is to keep only one parameter and $%
\Delta _{\text{SC}}$. We consider the case where electric field is applied
to a disk region of an antiferromagnetic topological superconductor ($%
\lambda _{\text{SX}}\neq 0$), as shown in Fig.\ref{pond}. Very strong
electric field can be applied experimentally by an STM\ probe. We assume
electric field is strong enough so that%
\begin{equation}
\lambda _{V}(r)=\ell E_{z}(r)=\pm \lambda _{\text{SO}}+\sqrt{\lambda _{\text{%
SX}}^{2}-\Delta _{\text{SC}}^{2}}.
\end{equation}%
The critical field is of the order of $E_{z}^{\text{cl}}=0.1V$\AA $^{-1}$.
The inner region of the circle have a nontrivial Chern number $C=1$ and
becomes a topological superconductor. On the other hands, the outer region
of the disk have $C=0$ and remains to be the trivial superconductor. As a
result, there emerges one Majorana fermion localized at the boundary of the
circle. Its wave function is given by (\ref{Circle}).

\end{document}